\documentclass[twocolumn,trackchanges]{aastex62}

    \usepackage{graphicx}	
    \usepackage{amsmath}	
    \usepackage{amssymb}	
    \usepackage{bm}
    \usepackage{longtable}
    \usepackage{color}

    \received{xxx}
    \revised{xxx}
    \accepted{xxx}
    \submitjournal{ApJ}
    
    \shorttitle{Cosmic Distance Duality Relation}
    \shortauthors{C.-Z. Ruan, F. Melia and T.-J. Zhang}
    
\begin{document}
    
    \title{Model-independent Test of the Cosmic Distance Duality Relation}

    \correspondingauthor{Fulvio Melia}
    
    \author{Cheng-Zong Ruan}
    \affiliation{Department of Astronomy, Beijing Normal University, \\
    Beijing 100875, China; chzruan@mail.bnu.edu.cn}
    
    \author{Fulvio Melia}
    \affiliation{Department of Physics, The Applied Math Program, and Department of Astronomy, \\
    The University of Arizona, AZ 85721, USA; fmelia@email.arizona.edu}
    
    \author{Tong-Jie Zhang}
    \affiliation{Department of Astronomy, Beijing Normal University, \\
    Beijing 100875, China; tjzhang@bnu.edu.cn}
    \affiliation{Institute for Astronomy Science, Dezhou University, \\
    Dezhou 253023, China}
    
    
    
    \begin{abstract}
    
    A validation of the cosmic distance duality (CDD) relation, $\eta(z)\equiv
    (1+z)^2d_A(z)/d_L(z)=1$, coupling the luminosity ($d_L$) and 
    angular-diameter ($d_A$) distances, is crucial because its violation 
    would require exotic new physics. We present a model-independent 
    test of the CDD, based on strong lensing and a reconstruction of the 
    HII galaxy Hubble diagram using Gaussian Processes, to confirm the 
    validity of the CDD at a very high level of confidence. Using parameterizations 
    $\eta(z) = 1 + \eta_0 z$ and $\eta(z) = 1 + \eta_1 z + \eta_2 z^2$, 
    our best-fit results are $\eta_0 = 0.0147^{+0.056}_{-0.066}$, and 
    $\eta_1 = 0.1091^{+0.1680}_{-0.1568}$ and $\eta_2 = -0.0603^{+0.0999}_{-0.0988}$, 
    respectively. In spite of these strong constraints, however, we also
    point out that the analysis of strong lensing using a simplified single
    isothermal sphere (SIS) model for the lens produces some irreducible scatter
    in the inferred CDD data. The use of an extended SIS approximation,
    with a power-law density structure, yields very similar results, but does not
    lessen the scatter due to its larger number of free parameters, which weakens
    the best-fit constraints. Future work with these strong lenses should
    therefore be based on more detailed ray-tracing calculations to 
    determine the mass distribution more precisely. 
    
    \end{abstract}
    
    
    
    \section{Introduction} \label{sec:intro}
    
    The cosmic distance duality (CDD) relation, based on 
    Etherington's theorem (1933), depends on three essential assumptions: 
    (i) that the spacetime is described by a metric theory of gravity; 
    (ii) that photons travel along null geodesics; and (iii) that their
    number is conserved along the null geodesics. The CDD is commonly
    written in the form $\eta(z) = 1$, with the definition
    \begin{align}
        \eta(z) \equiv (1+z)^2 \frac{d_A(z)}{d_L(z)} \ , \label{equ:cddeta}
    \end{align}
    where $d_A(z)$ and $d_L(z)$ are the angular-diameter and luminosity distances, 
    respectively.
    
    Many attempts have been made to test the validity of the CDD, using several
    different kinds of data, and/or assumptions. Typically, the angular-diameter 
    distance $d_A(z)$ is measured using the angular size of galaxy clusters
    \citep{wei2015clusters,melia2016clusters}, while 
    the luminosity distance $d_L(z)$ is often inferred from Type Ia SNe. For a 
    non-exhaustive set of references, see \citet{2004PhRvD..69j1305B}; 
\citet{uzan2004distance}; \citet{holanda2010testing}, \citet{holanda2012probing}; \citet{khedekar2011new}; \citet{li2011cosmological}; \citet{nair2011observational}; \citet{lima2011deformed}; \citet{meng2012morphology}; \citet{ellis2013blackness}; \citet{liao2016distance}; \cite{2017arXiv171010929Y}; \citet{hu2018testing}; \citet{melia2018model}. 
    But a principal difficulty with using SNe is that the measurement of $d_L(z)$ 
    is model-dependent. One can easily see this from the definition of the distance 
    modulus $\mu$, which is given as
    \begin{align}
        \mu = 5\log d_L - 5 = m_{\mathrm{max}} - M_{\mathrm{max}} \ ,
    \end{align}
    in terms of the peak magnitude $m_{\mathrm{max}}$ and peak absolute magnitude
    $M_{\mathrm{max}}$. Every Type Ia SN has almost the same $M_{\mathrm{max}}$,
    so if $m_{\mathrm{max}}$ is measured, one can obtain the distance modulus $\mu$. 
    The difficulty arises from the scatter in peak magnitudes, which depend rather
    strongly on the shapes and colors of the SN lightcurves \citep{guy2005salt}. 
    To get $m_{\mathrm{max}}$, one of several fitters must be used to parameterize the light 
    curves. For example, one of the most popular parameterizations is \citep{guy2007salt2}
    \begin{align}
        \mu_B (\alpha, \beta, M_B; z) = m_B^{\mathrm{max}} (z) - M_B + \alpha x - \beta c \ ,
    \end{align}
    where $m_B^{\mathrm{max}}$ is the rest-frame peak magnitude of the $B$ band, $x$ is 
    the stretch factor that describes the effect of the lightcurve shape on $\mu$, 
    and $c$ is the color parameter representing the influence of intrinsic color and 
    reddening due to dust on $\mu$. The so-called `nuisance' parameters $\alpha$, 
    $\beta$, and $M_B$ must be optimized along with all the other parameters in the
    chosen cosmological model. By now, it is well known that different models are
    associated with different values of these nuisance variables, so there is no
    unique way to determine the SN distance moduli in a truly model independent way.
    It is therefore quite likely that some (or all) of the previously claimed CDD 
    violations may simply be due to unaccounted for influences of the assumed 
    cosmology on $\eta(z)$ \citep{uzan2004distance,holanda2010testing,li2011cosmological}. 
    For a more detailed discussion, see 
   \citet{yang2013improved,melia2009,melia2012SNe,melia2013,wei2015comparative,melia2018model}. 
    
    In this paper, we steer clear of measurements that require the pre-assumption of
    particular cosmological models, and instead use strong lenses to measure the 
    ratio of angular-diameter distances, and a reconstruction of the HII galaxy Hubble 
    diagram with Gaussian Processes to obtain the luminosity distance. In the next
    section, we shall discuss the rationale behind these two kinds of observation, 
    and why one may safely assume model independence in the associated data.
    Since no model is assumed in any of our analysis, our approach yields a clean 
    measure of the CDD relation.
    
    We shall first briefly summarize the methodology of measuring $\eta(z)$ as a
    function of redshift using strong lensing and HII galaxies in \S~\ref{sec:methd}. 
    We then describe the relevant datasets in \S~\ref{sec:datasets}, and present 
    the results of our analysis in \S~\ref{sec:res}. We shall demonstrate that
    this combination of observations confirms the CDD at a very high level of 
    confidence. Finally, we present our conclusions in \S~\ref{sec:conc}.

    \section{Methodology}\label{sec:methd}
    
    \subsection{Angular-diameter distance from strong lensing}
    An Einstein ring is formed when the source, lens, and observer are aligned along 
    the same line of sight. For a strong lensing system with a single galaxy as a 
    lens, the Einstein ring's radius $\theta_E$ depends on the ratio of 
    angular-diameter distances between the lens and source, $d_A(z_l, z_s)$, 
    and the observer and source, $d_A(0, z_s)$, and the mass distribution within 
    the lensing galaxy. The lens galaxy model is sometimes simplified as 
    a single isothermal ellipsoid (SIE) \citep{ratnatunga1999top, kochanek2000fundamental}. 
    Several prior analyses have shown that a reasonable further simplification may
    be adequate, in which the ellipsoid is replaced by a sphere (SIS) 
    \citep{1538-4357-602-1-L5,2015ApJ...806..185C,melia2014comparison}, i.e., with
    zero ellipticity. To keep the analysis as straightforward as possible, we also
    adopt this approach for our study in this paper. To gauge the impact
    of this approximation, however, we shall also compare our results to those obtained
    with an extended SIE lens model (see \S~4.2 below), in which the mass distribution is 
    assumed to be a power-law with optimizable parameters 
    \citep{koopmans2005proc,doi:10.1093/mnras/stw932,1538-3881-121-4-1936,0004-637X-834-1-75}.
    
    With the SIS mass distribution, 
    the Einstein radius is expressed as \citep{schneider2006gravitational,rana2017probing}
    \begin{align}
        \theta_E = 4\pi \frac{d_A(z_l, z_s)}{d_A(0, z_s)} \left( \frac{\sigma_{\mathrm{SIS}}}{c} \right)^2 \ , \label{equ:sl1}
    \end{align}
    where $c$ is the speed of light and $\sigma_{\mathrm{SIS}}$ is the velocity 
    dispersion of the lens mass distribution. Therefore, the quantity that comes
    directly from observation of a strong lensing system is the distance ratio 
    \begin{align}
        d_{\mathrm{ratio}} &\equiv \frac{d_A(z_l, z_s)}{d_A(0, z_s)} = \frac{\theta_E}{4\pi} \left( \frac{c}{\sigma_{\mathrm{SIS}}} \right)^2 \ . \label{equ:dratiodef}
    \end{align}
    
    In a spatially flat cosmology, one may write for the comoving distance 
    \begin{equation}
    r(z_l, z_s) = r(z_s) - r(z_l)\;.
    \end{equation}
    Thus, using 
    \begin{equation}
    d_A(z) = r(z) / (1+z)\;,
    \end{equation}
    the angular-diameter distance between lens and source, $d_A(z_l, z_s)$, may
    be expressed in terms of $d_A(z_l)$ and $d_A(z_s)$:
    \begin{align}
        d_{\mathrm{ratio}} = 1 - \left(\frac{1+z_l}{1+z_s}\right) 
        \frac{d_A (z_l)}{d_A(z_s)} \ . \label{equ:dratio1}
    \end{align}
    Note, however, that for a non-flat cosmology, the comoving distance depends on
    $\sin$- or $\sinh$-like functions \citep{PhysRevLett.115.101301}, so a simple 
    relation like Eq. \eqref{equ:dratio1} is not possible. Fortunately, most 
    cosmological observations today favor a flat universe \citep{refId0}, so
    the spatially flat assumption is not a stringent restriction for our CDD test.
    
\begin{center}
\begin{longtable}{lllllc}
\caption{Strong-lensing systems used in this paper, sorted \\ by source redshift $z_s$}\\
\hline\hline
Name &  $z_l$ &  $z_s$ &  $\theta_E$ &Survey&$\sigma_0$\\
&&&($''$)&&(km s$^{-1}$)\\
\hline
&&\\
    J0219-0829 & 0.389 & 2.15  & 1.3   & SL2S  & 298$\pm$24 \\
    J0849-0251 & 0.274 & 2.09  & 1.16  & SL2S  & 279$\pm$35 \\
    J0214-0405 & 0.609 & 1.88  & 1.41  & SL2S  & 293$\pm$48 \\
    J0217-0513 & 0.646 & 1.847 & 1.27  & SL2S  & 253$\pm$29 \\
    J1404+5200 & 0.456 & 1.59  & 2.55  & SL2S  & 342$\pm$20 \\
    J0849-0412 & 0.722 & 1.54  & 1.1   & SL2S  & 338$\pm$25 \\
    J1406+5226 & 0.716 & 1.47  & 0.94  & SL2S  & 262$\pm$20 \\
    J2122+0409 & 0.626 & 1.452 & 1.58  & BELLS & 326$\pm$56 \\
    J0223-0534 & 0.499 & 1.44  & 1.22  & SL2S  & 293$\pm$28 \\
    J0830+5116 & 0.53  & 1.332 & 1.14  & BELLS & 274$\pm$37 \\
    J1215+0047 & 0.642 & 1.297 & 1.37  & BELLS & 266$\pm$46 \\
    J0226-0420 & 0.494 & 1.232 & 1.19  & SL2S  & 272$\pm$25 \\
    J1631+1854 & 0.408 & 1.086 & 1.63  & BELLS & 272$\pm$14 \\
    J1420+5258 & 0.38  & 0.99  & 0.96  & SL2S  & 252$\pm$24 \\
    J2125+0411 & 0.363 & 0.978 & 1.2   & BELLS & 250$\pm$17 \\
    J2303+0037 & 0.458 & 0.936 & 1.02  & BELLS & 278$\pm$31 \\
    J0157-0056 & 0.513 & 0.924 & 0.79  & SLACS & 308$\pm$49 \\
    J0747+5055 & 0.438 & 0.898 & 0.75  & BELLS & 323$\pm$59 \\
    J0747+4448 & 0.437 & 0.897 & 0.61  & BELLS & 286$\pm$53 \\
    J1250+0523 & 0.232 & 0.795 & 1.13  & SLACS & 257$\pm$14 \\
    J1630+4520 & 0.248 & 0.793 & 1.78  & SLACS & 281$\pm$16 \\
    J1531-0105 & 0.16  & 0.744 & 1.71  & SLACS & 281$\pm$14 \\
    J1525+3327 & 0.358 & 0.717 & 1.31  & SLACS & 264$\pm$26 \\
    J1213+6708 & 0.123 & 0.64  & 1.42  & SLACS & 291$\pm$15 \\
    J0037-0942 & 0.196 & 0.632 & 1.53  & SLACS & 283$\pm$10 \\
    J1204+0358 & 0.164 & 0.631 & 1.31  & SLACS & 275$\pm$17 \\
    J1112+0826 & 0.273 & 0.63  & 1.49  & SLACS & 329$\pm$21 \\
    J0324-0110 & 0.4456 & 0.6239 & 0.63  & SLACS2017 & 314$\pm$38 \\
    J0946+1006 & 0.222 & 0.608 & 1.38  & SLACS & 266$\pm$21 \\
    J0822+2652 & 0.241 & 0.594 & 1.17  & SLACS & 264$\pm$15 \\
    J0737+3216 & 0.322 & 0.581 & 1     & SLACS & 339$\pm$17 \\
    J1430+4105 & 0.285 & 0.575 & 1.52  & SLACS & 324$\pm$32 \\
    J1020+1122 & 0.282 & 0.553 & 1.2   & SLACS & 289$\pm$18 \\
    J0109+1500 & 0.294 & 0.525 & 0.69  & SLACS & 259$\pm$20 \\
    J0216-0813 & 0.332 & 0.524 & 1.16  & SLACS & 335$\pm$23 \\
    J1627-0053 & 0.208 & 0.524 & 1.23  & SLACS & 295$\pm$14 \\
    J2303+1422 & 0.155 & 0.517 & 1.62  & SLACS & 254$\pm$16 \\
    J1101+1523 & 0.178 & 0.5169 & 1.18  & SLACS2017 & 283$\pm$16 \\
    J1205+4910 & 0.215 & 0.481 & 1.22  & SLACS & 283$\pm$14 \\
    J1402+6321 & 0.205 & 0.481 & 1.35  & SLACS & 268$\pm$17 \\
    J0956+5100 & 0.24  & 0.47  & 1.33  & SLACS & 338$\pm$17 \\
    J0955+3014 & 0.3214 & 0.4671 & 0.54  & SLACS2017 & 271$\pm$33 \\
    J0935-0003 & 0.348 & 0.467 & 0.87  & SLACS & 391$\pm$35 \\
    J2300+0022 & 0.228 & 0.464 & 1.24  & SLACS & 285$\pm$17 \\
    J1016+3859 & 0.168 & 0.439 & 1.09  & SLACS & 254$\pm$13 \\
    J1106+5228 & 0.096 & 0.407 & 1.23  & SLACS & 268$\pm$13 \\
    J1143-0144 & 0.106 & 0.402 & 1.68  & SLACS & 264$\pm$13 \\
    J1543+2202 & 0.2681 & 0.3966 & 0.78  & SLACS2017 & 288$\pm$16 \\
    J0920+3028 & 0.2881 & 0.3918 & 0.7   & SLACS2017 & 293$\pm$17 \\
    J1330+1750 & 0.2074 & 0.3717 & 1.01  & SLACS2017 & 251$\pm$12 \\
    J0912+0029 & 0.164 & 0.324 & 1.63  & SLACS & 323$\pm$12 \\
    J2324+0105 & 0.1899 & 0.2775 & 0.59  & SLACS2017 & 255$\pm$16 \\
    J0044+0113 & 0.12  & 0.196 & 0.79  & SLACS & 267$\pm$13 \\
&& \\
\hline\hline
\end{longtable}
\end{center}

    \subsection{Luminosity distance from GP reconstruction of the HII galaxy Hubble diagram}
    The hydrogen gas ionized by massive star clusters in HII galaxies emits prominent 
    Balmer lines in $\mathrm{H}\alpha$ and $\mathrm{H}\beta$ 
    \citep{terlevich1981dynamics,kunth2000most}. The luminosity $L(\mathrm{H}\beta)$ 
    in $\mathrm{H}\beta$ from these structures is strongly correlated with the velocity 
    dispersion $\sigma_v$ of the ionized gas \citep{terlevich1981dynamics}, because 
    both the intensity of ionizing radiation and $\sigma_v$ increase with the starbust 
    mass \citep{siegel2005towards}. The relatively small dispersion in the relationship 
    between $L(\mathrm{H}\beta)$ and $\sigma_v$ allows these galaxies and local HII 
    regions to be used as standard candles \citep{terlevich2015road,wei2016h,leafmelia2018a}.
    
    The luminosity of HII galaxies versus their velocity dispersion correlation is 
    \citep{terlevich2015road} 
    \begin{align}
        \log L(\mathrm{H}\beta) = \alpha \log \sigma_v (\mathrm{H}\beta) + \kappa \ , \label{eqn:Lsigmarelation}
    \end{align}
    where $\alpha$ and $\kappa$ are constants. As was the case with Type Ia SNe, 
    these two parameters in principle need to be optimized simultaneously with 
    those of the cosmological model. \citet{wei2016h} have shown, however, that 
    their values are insensitive to the adopted model, and appear to be universal.
    This is the important step that allows us to use the HII galaxy Hubble diagram
    in a model-independent way. For example, comparing the two distinct cosmologies
    $R_{\rm h}=ct$ and $\Lambda$CDM, and defining an `$H_0 \text{-} \text{free}$' 
    logarithmic lumiosity parameter 
    \begin{align}
        \delta \equiv -2.5\kappa - 5 \log \left( \frac{H_0}{\mathrm{km}\;
        \mathrm{s}^{-1}\;\mathrm{Mpc}^{-1}} \right)  + 125.2 \ .
    \end{align}
    \citet{wei2016h} showed that $\alpha = 4.86^{+0.08}_{-0.07}$ and 
    $\delta = 32.38^{+0.29}_{-0.29}$ for the former, while $\alpha = 4.89^{+0.09}_{-0.09}$ 
    and $\delta = 32.49^{+0.35}_{-0.35}$ for the latter. Such small differences fall
    well within the observational uncertainty (note, e.g., that the difference in 
    $\alpha$ is $\sim \sigma/3$). We shall therefore simply use the average values 
    of these `nuisance' parameters, i.e., $\alpha = 4.87^{+0.11}_{-0.08}$ and 
    $\delta = 32.42^{+0.42}_{-0.33}$. The distance modulus of an HII galaxy is
    \begin{equation}
        \mu^{\mathrm{obs}} (z) = -\delta + 2.5 \left[ \alpha \log \sigma_v (\mathrm{H} \beta) 
        - \log F(\mathrm{H}\beta) \right] \ ,
    \end{equation} 
    and the luminosity distance is correspondingly
    \begin{align}
        d_L^{\mathrm{obs}} (z) = 10^{(\mu^{\mathrm{obs}} (z)/5 - 5)} \;\mathrm{Mpc} \ .\label{equ:dlum}
    \end{align}

    For every $d_A(z_i)$ measurement from a strong-lensing system, we use a 
    model-independent Gaussian Process (GP) reconstruction to get the 
    corresponding $d_L(z_i)$. A description of the GP approach and its
    use with the HII galaxy Hubble diagram may be found in 
    \citet{yennapureddy2017reconstruction,yennapureddy2018cosmological}, 
    based on the pioneering work of \citet{seikel2012reconstruction}.

    It is important to point out an important caveat, however, having to do with
    possible systematic uncertainties in the HII galaxy probe, specifically the 
   $L(\mathrm{H}\beta) \text{-} \sigma$ correlation, which still needs to be fully understood. 
    Systematic uncertainties in this critical relation include the size of the starburst, its age, the 
   oxygen abundance of HII galaxies and the internal extinction correction \citep{doi:10.1093/mnras/stw1813}. 
    The scatter found in the $L(\mathrm{H}\beta) \text{-} \sigma$ relation 
   (Equation ~\eqref{eqn:Lsigmarelation}) for HII galaxies suggests that there may exit a 
   possible dependence on a second parameter. Some progress has already been made in trying 
   to mitigate these uncertainties. For example, \citet{doi:10.1093/mnras/stu987} find that for a 
   sample of local HII galaxies, the size of the star-forming region can serve as the second 
   parameter.  

   Another important consideration is the exclusion of the rotating support for the 
   system (as opposed to purely kinematic support), which would obviously distort the 
   $L(\mathrm{H}\beta) \text{-} \sigma$ relation. 
   \citet{doi:10.1093/mnras/stu987,doi:10.1093/mnras/stw1813} have suggested using an 
   upper limit of  the velocity dispersion to minimize this possibility, although the catalogue of 
   suitable sources is then greatly reduced. However, even with this limitation, there is no 
   guarantee that this systemic effect will be completely eliminated. Our results in this paper
   should be viewed with this cautionary consideration, with the hope and expectation that
   future improvements in our understanding of these systems will make the HII Hubble
   diagram an even more powerful probe of the integrated distance measure than it is now.
    
    \subsection{Measurement of the CDD relation}
   As one can see from Equations \eqref{equ:dratiodef} and \eqref{equ:dratio1}, 
the lensing measurements provide the values of $\theta_E$ and $\sigma_\mathrm{SIS}$, 
which give the ratio of angular-diameter distances to $z_l$ and $z_s$, i.e. $d_A(z_l)/d_A(z_s)$. 
For each lens and each source, we then form the ratio given in Equation \eqref{equ:cddeta} by 
calculating the luminosity distance $d_L(z_l)$ and $d_L(z_s)$ from the GP reconstruction of the 
HII Hubble diagram:
    \begin{align}
        \frac{\eta(z_l)}{\eta(z_s)} &= \frac{(1+z_l)^2}{(1+z_s)^2} \frac{d_A (z_l)}{d_A(z_s)} \frac{d_L(z_s)}{d_L(z_l)} \notag \\
         &= 
        10^{\left[\mu^{\mathrm{obs}}(z_s) - \mu^{\mathrm{obs}}(z_l)\right]/5 }
        \left(\frac{1+z_l}{1+z_s}\right)\times\nonumber \\ 
        \null &\qquad\qquad
        \left(1-{\theta_E\over 4\pi}\left[{c\over \sigma_{\mathrm{SIS}}}\right]^2\right). \label{etaratio2}
    \end{align}
    If the CDD relation is realized in nature, this ratio should always
    be equal to $1$, independently of $z_l$ and $z_s$. Our approach uses the
    strong-lensing and HII galaxy data to `measure' this ratio and test
    the CDD hypothesis for a broad range of redshift pairs $(z_l,z_s)$.

    \section{Data}\label{sec:datasets}
    We extract our strong lenses from the catalog of 158 sources recently compiled by
    \citet{doi:10.1093/mnras/sty1365} (see Table 1 therein.) from the SLACS, 
BELLS, LSD and SL2S surveys. We keep only strong-lensing systems with $\sigma_0 \ge 250\,
\mathrm{km}\,\mathrm{s}^{-1}$ and $z_s \le 2.33$, which leaves $53$ strong lenses 
(see Table~1 below). The former condition mitigates the scatter 
when using the SIS model, while the latter condition is imposed by the limit of the highest redshift 
in the HII galaxy data, which is $2.33$. The uncertainty in the observed value of $d_{\rm ratio}$ 
is estimated from the measurement error of the Einstein radius, $\sigma_{\theta_E}$, and the velocity 
dispersion, $\sigma_{\sigma_0}$. Using standard error propagation, the 
corresponding dispersion for $d_{\mathrm{ratio}}$ is
    \begin{align}
        \sigma_{d_{\mathrm{ratio}}} = d_{\mathrm{ratio}} \sqrt{ \left(\frac{\sigma_{\theta_E}}{\theta_E}\right)^2 + 4\left(\frac{\sigma_{\sigma_0} }{\sigma_0}\right)^2 } \ ,
    \end{align}
    in which we assume a uniform $5\%$ error for $\sigma_{\theta_E}$, following 
    \citet{grillo2008cosmological}. There is also the question of how the dispersion
    $\sigma_{\rm SIS}$ for a SIS model relates to $\sigma_0$. Previous work, e.g., 
    by \citet{melia2014comparison} and \citet{2015ApJ...806..185C}, indicates that the 
    statistics of a large sample of lenses is consistent with a SIS model for the 
    lens mass distribution with $\sigma_{\mathrm{SIS}} = \sigma_0$. Fits to the data
    with a more general relation, $\sigma_{\mathrm{SIS}} = f_e \sigma_0$, with $f_e$
    a parameter to be optimized, shows that its optimial value is within only a few
    percentage points of $1$ \citep{liao2016distance}.
    
    Nonetheless, recent studies have also shown that a pure SIS model may not be
    an accurate representation of the lens mass distribution when 
    $\sigma_0 < 250 \;\mathrm{km}\;\mathrm{s}^{-1}$ \citep{doi:10.1093/mnras/sty1365} 
    (see Figure 2 therein), for which unphysical values of $d_{\rm ratio}$ are often
    encountered. \citet{doi:10.1093/mnras/sty1365} found  that ignoring individual 
    variations from a pure SIS structure results in unsatisfactory fitting results. 
    For example, a flattened lens galaxy distribution, corresponding to a small 
    $\sigma_0$, deviates significantly from a pure SIS model. To avoid systematics 
    such as this, we shall therefore select from the overall lens sample only those 
    sources with $\sigma_0 \ge 250 \;\mathrm{km}\;\mathrm{s}^{-1}$.  

    \begin{figure}
        \plotone{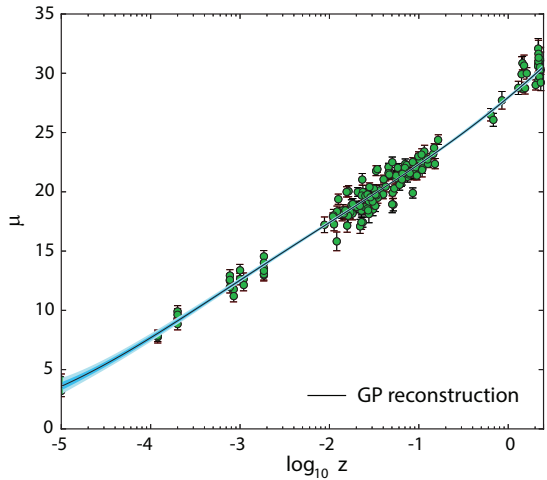}
        \caption{Distance modulus of the currently available HII-region/Galaxy observations, shown with $1\sigma$ error bars, spanning a redshift range up to $\sim 2.33$. The GP reconstructed distance modulus $\mu(z)$ is shown as a solid black curve, with its $1\sigma$
confidence region (shaded swath). (Adapted from \citet{yennapureddy2017reconstruction}). The data sample consists of $25$ high-$z$ HII galaxies, $107$ local HII galaxies, and $24$ giant extragalactic HII regions, for a total of $156$ sources \citep{terlevich2015road}. \label{fig:fig1}}
    \end{figure}
        
    For the HII galaxy Hubble diagram, we use the $25$ high-$z$ HII galaxies, $107$ 
    local HII galaxies, and $24$ giant extra galactic HII regions from the catalog
    compiled by \citet{terlevich2015road}. The GP reconstructed distance modulus, 
    $\{z,\;\mu^{\mathrm{obs}}(z),\;\sigma_{\mu^{\mathrm{obs}}}\}$, is calculated
    from these data following the prescription described and implemented in 
    \citet{yennapureddy2017reconstruction}. The distance modulus data and GP 
reconstruction are shown in figure \ref{fig:fig1}.  As one can see in this figure, the 
error in the reconstructed curve is smaller than that of the individual data points. As discussed 
more extensively in \citet{yennapureddy2017reconstruction}, the confidence region depends 
on the errors in the data, $\sigma_{\mu_{\mathrm{obs}}}$, on the optimized hyperparameter(s) 
of the GP method---such as the characteristic `bumpiness' parameter $\sigma_f$---and on the 
product of the covariance matrixes $K_* K^{-1} K_*^T$ between the estimation points and 
dataset points (see \citet{yennapureddy2017reconstruction}). The reconstructed uncertainty 
$\sigma_{\mu_{\mathrm{GP}}} (z_*)$ at the estimation point $z_*$ is less than 
$\sigma_{\mu_{\mathrm{obs}}}$ when, for the point $z_*$, there is a large correlation 
between the data, $K_* K^{-1} K_*^T > \sigma_f$, which is most of the time for the HII 
galaxy data used in this study. As such, the estimated confidence region is smaller than 
that of the observational data.
    
    The associated $\eta$-ratio errors are 
    estimated from Equation \eqref{etaratio2} using standard error propagation.
    Defining 
    \begin{align}
        k & \equiv 10^{[\mu^{\mathrm{obs}}(z_l) - \mu^{\mathrm{obs}}(z_s)]/5} \ , \\
        g & \equiv \frac{1+z_l}{1+z_s} (1 - d_{\mathrm{ratio}}) \ ,
    \end{align}
    we have
    \begin{align} 
        \sigma_k &= k\times \frac{\ln 10}{5} \sqrt{ \sigma_{\mu^{\mathrm{obs}}(z_l)}^2 
        + \sigma_{\mu^{\mathrm{obs}}(z_s)}^2 } \ , \\
        \sigma_g &= \frac{1+z_l}{1+z_s} \sigma_{d_{\mathrm{ratio}}} \ , 
    \end{align}
    so that
    \begin{equation}
        \sigma_{\eta(z_l)/\eta(z_s)} = \frac{\eta(z_l)}{\eta(z_s)} \sqrt{ \left(\frac{\sigma_k}{k}\right)^2 + \left(\frac{\sigma_g}{g}\right)^2 } \ . \label{eqn:sigmaeta}
    \end{equation}
   
\vskip 0.2in
    \section{Results and Discussion}
    \label{sec:res}
    \subsection{SIS Model}
\begin{figure}
          \plotone{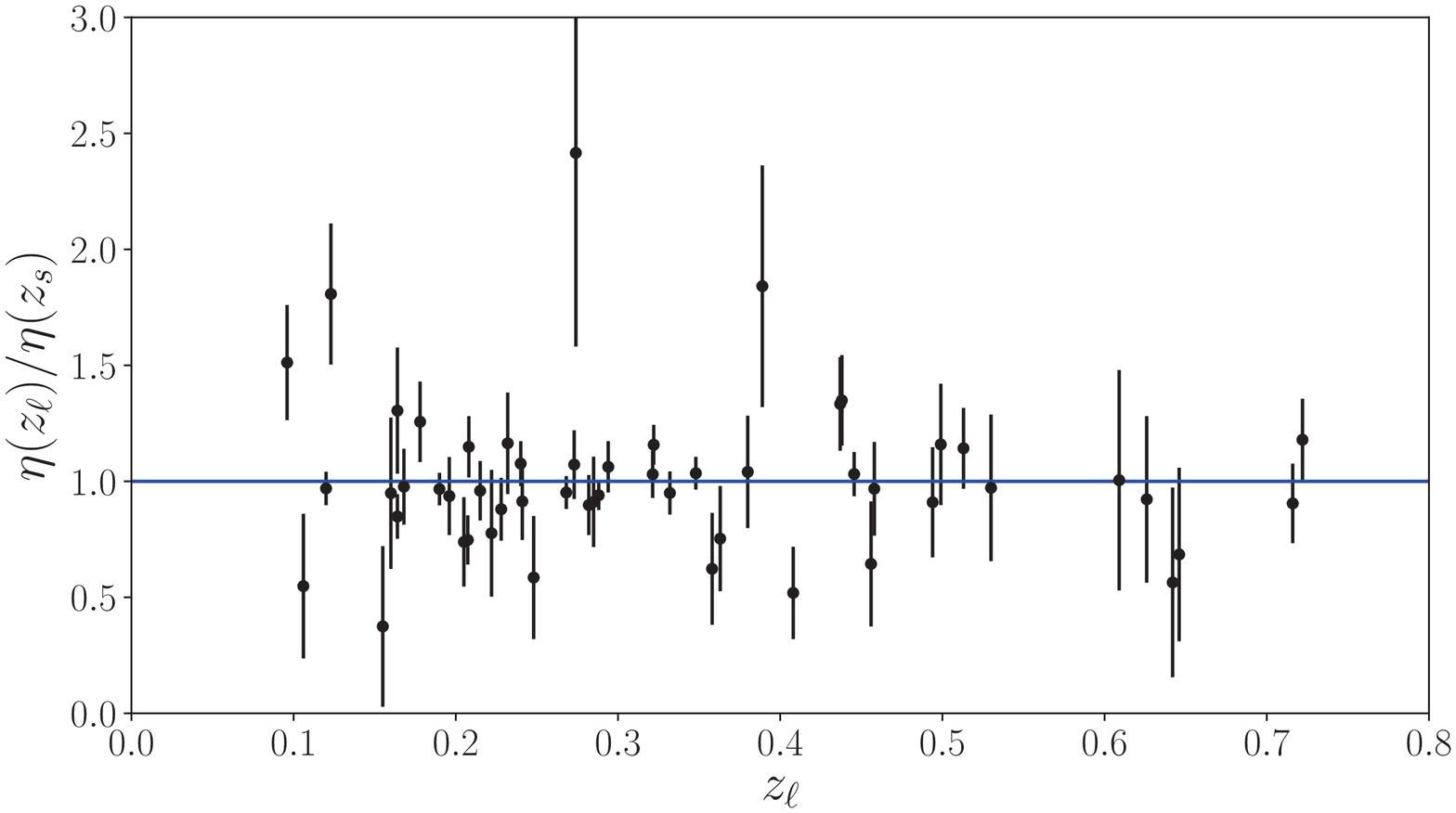}
          \caption{Measured values of the ratio $\eta(z_l)/\eta(z_s)$ versus $z_\ell$, estimated 
          from strong lensing systems and the GP reconstruction of the HII galaxy Hubble diagram. 
          These data confirm the CDD at a very high level of confidence. \label{fig:fig2}}
      \end{figure}

\begin{figure}
        \plotone{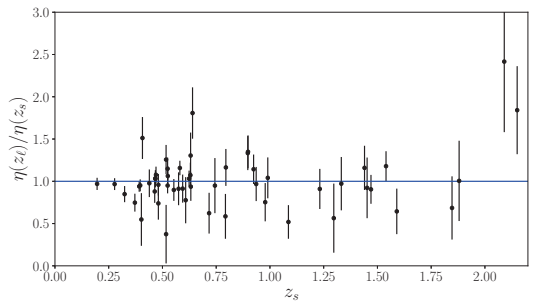}
        \caption{Measured values of the ratio $\eta(z_l)/\eta(z_s)$ versus $z_s$, estimated 
        from strong lensing systems and the GP reconstruction of the HII galaxy Hubble diagram. 
        These data confirm the CDD at a very high level of confidence. \label{fig:fig3}}
    \end{figure}

    Let us first analyze these data using the simplest SIS model 
    described in \S~2.1 above. We shall estimate the impact of changing the mass
    distribution in the lens by considering an extension to this profile in 
    \S4.2. The 53 $\{z_l, z_s, \eta(z_l)/\eta(z_s) \}$ data points obtained with 
    the use of Equation \eqref{etaratio2} and the data described in \S~3 are 
    plotted as a function of $z_l$ in Figure \ref{fig:fig2}, and as a function
    of $z_s$ in Figure \ref{fig:fig3}. As one can see from these distributions, 
    our test of the CDD extends over a significantly large redshift range, with 
    several sources beyond $z_s\sim 2$. 
    
    To gauge whether these data confirm or reject the CDD, we parameterize 
    $\eta(z)$ using the following two forms:
    \begin{align}
        \eta (z) &= 1 + \eta_0 z \ , \label{equ:etapara1} \\
        \eta (z) &= 1 + \eta_1 z + \eta_2 z^2 \ , \label{equ:etapara2}
    \end{align}
    in which
    $\eta_0,\;\eta_1$ and $\eta_2$ are all assumed to be constant. The CDD 
    corresponds to $\eta(z)\equiv 1$, i.e., $\eta_0 = \eta_1 = \eta_2 = 0$. 
  
    To find the best-fitting CDD parameters and the confidence regions, we use 
Bayesian statistical methods and the Markov chain Monte Carlo (MCMC) technique to calculate 
the posterior probability density function (PDF) of the parameter matrix $\bm{\eta} \equiv 
\{\eta_0 \} \ \mathrm{or}\  \{\eta_1, \eta_2\}$, which is
    \begin{align}
        p(\bm{\eta} | \text{eta ratio data}) \propto \mathcal{L}(\bm{\eta}, \text{eta ratio  data}) 
        \times p(\bm{\eta}) \ ,
    \end{align}
    where
    \begin{itemize}
    \item $p(\bm{\eta})$ is the prior and (assumed) uniform distribution;
    \item $\mathcal{L} \propto \exp (-\chi^2/2)$ is the likelihood function, and 
    \begin{align}
    \chi^2 = \sum_i \left\{ \frac{\left[ \eta(z_{l, i})/\eta(z_{s,i})\right]^{\text{obs}} 
    - \eta(\bm{\eta}; z_l)/\eta(\bm{\eta}; z_s)  }{\sigma_{\eta(z_{l, i})/\eta(z_{s,i})}} \right\}^2
    \end{align}
    is the $\chi^2$ function, with $\eta(\bm{\eta}; z) \equiv 1+\eta_0 z$ or $1 + 
    \eta_1 z + \eta_2 z^2$, as the case may be.
    \end{itemize}
    
\begin{figure}
    \plotone{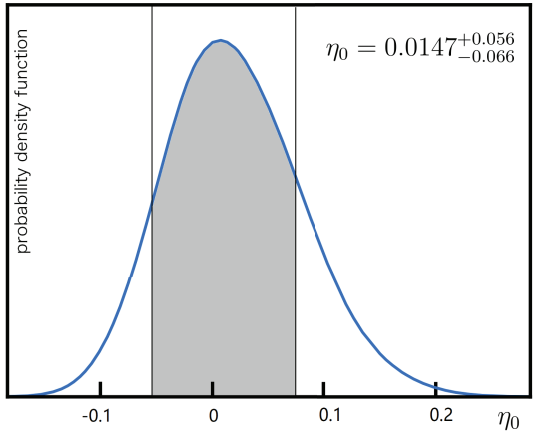}
    \caption{Posterior probability density function of the parameter $\eta_0$ in
    the CDD parameterization $\eta(z)=1+\eta_0 z$. Its optimized value is consistent
    with $0$ to within $\sim \sigma/4$.\label{fig:fig4}}
    \end{figure}

\begin{figure}
    \plotone{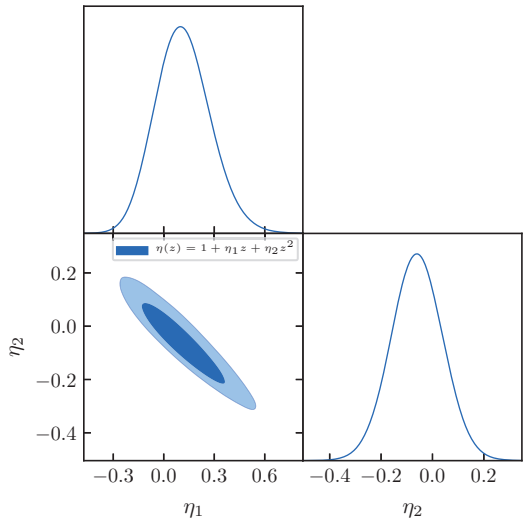}
    \caption{Posterior probability density function of the parameters $\eta_1$ and $\eta_2$
    in the CDD parameterization $\eta(z)=1+\eta_1 z + \eta_2 z^2$. Both of the
    optimized values are consistent with $0$ at better than $\sim 2\sigma/3$.\label{fig:fig5}}
    \end{figure}

    The MCMC method uses the Metropolis-Hastings algorithm to generate Markov 
chains of sample points in parameter space from the posterior probability \citep{2014sdmm.book.....I}. 
We use the emcee Python module\footnote{\url{http://dfm.io/emcee/current/}}
    \citep{2013PASP..125..306F} that implements Markov chain Monte Carlo to sample 
    from the posterior distribution of parameters ($\eta_0$, or $\eta_1$ and 
    $\eta_2$). For the two sets of data used in this analysis, we find that
    the optimized values of the parameters in $\eta(z)$, and their $1\sigma$ 
    errors, are 
    \begin{align}
        \eta_0 &= 0.0147^{+0.056}_{-0.066} \ , \label{equ:eta0SIS}\\
        \eta_1 &= 0.1091^{+0.1680}_{-0.1568}\ , \quad \eta_2 = -0.0603^{+0.0999}_{-0.0988} \ .
    \end{align}
    The PDF plots are shown in Figure \ref{fig:fig4} and \ref{fig:fig5}. 
    The contours were plotted using the Python package 
    ``GetDist''\footnote{\url{http://getdist.readthedocs.io/en/latest/}}. 
    As one can see, the best-fit values of both $\eta_0$, and $\eta_1$ and $\eta_2$,
    are entirely consistent with the CDD at a very high level of confidence. Specifically,
    the strong lensing data, in combination with the HII galaxy Hubble diagram,
    show that $\eta_0$ deviates from zero by less than $\sim \sigma/4$, while
    $\eta_1$ and $\eta_2$ deviate from zero by less than $\sim 2\sigma/3$.
    
    Notice, however, that the parameterization in Equation~\eqref{equ:etapara1} appears to give
   a more accurate result than that in Equation~\eqref{equ:etapara2}. A quick inspection of 
   Figures \ref{fig:fig2} and \ref{fig:fig3} 
    suggests why the simpler parameterization in the former appears to confirm the 
    CDD more strongly than that in the latter.
    Though the $\eta(z_l)/\eta(z_s)$ data points fluctuate across $\eta(z)=1$ throughout
    the redshift range of interest, some points clearly deviate from this value by
    $1\sigma-2\;\sigma$. This scatter produces enhanced fluctuation in the redshift
    dependence of the parameterized $\eta(z)$ function when a higher-order polynomial
    is used, which accounts for the fact that both $\eta_1$ and $\eta_2$ differ from
    zero by a larger fraction of their $\sigma$ than does $\eta_0$. 
     
    Unfortunately, this residual scatter is due to the over-simplifying assumption that all 
    lens mass distributions follow the SIS, as noted earlier by \citet{doi:10.1093/mnras/sty1365}. 
   For example, all lens systems with $d_A(z_l) > d_A (z_s)$---which is clearly 
   unphysical---have a small velocity dispersion $\sigma_0 \le 233\,\mathrm{km}\,
   \mathrm{s}^{-1}$. And, as shown in Figure 2 of \citet{doi:10.1093/mnras/sty1365}, 
   there is an obvious correlation between the velocity dispersion $\sigma_0$ and 
   $d_{\mathrm{ratio}}$, i.e., a correlation between the mass distribution of the lens 
   galaxy and the lens and source's distance ratio, which cannot be accounted for
   in the SIS model. This significant correlation is a hint that the SIS approximation
   is reliable only for strong lensed systems.
    We have largely mitigated this effect by restricting our analysis to systems with
    $\sigma_0\ge 250$ km s$^{-1}$, but variations away from the pure SIS model are
    apparently present even for this reduced sample. Our results already provide
    compelling evidence that the CDD is realized in nature, but we suggest that a
    parameterization such as that in Equation \eqref{equ:etapara2} can do even better in the future
    if the lens mass distribution were to be determined more accurately (say, with 
    ray tracing), rather than the adoption of a simple SIS model.

    Of course, the uncertainties of the HII galaxy data that provide the luminosity 
    distance also contribute to the $\eta$-ratio's scatter. But as one can see from 
    Figure~\ref{fig:fig1} and the discussion of the GP reconstruction in 
    Section~\ref{sec:datasets}, the relative uncertainty in the GP reconstructed HII galaxy distance modulus is much smaller than that of the strong lensing data.  The dominant contribution to the error in Equation ~\eqref{eqn:sigmaeta} therefore comes from the 
    lens data rather than the HII galaxy data.

    \subsection{Extended SIE Model}
    As noted earlier, we can explore this hypothesis further by examining how 
    a change in the lens model affects the scatter seen in figs.~2 and 3. To do so, however,
    we need to rexamine how one converts the observable quantities of strong lensing systems,
    such as the Einstein radius $\theta_E$ and the central velocity dispersion $\sigma$ of the 
    lensing galaxy, into the angular-diameter distance. For the SIS model, the relation between 
    these observables and $d_A$ is Equation \eqref{equ:sl1} or \eqref{equ:dratiodef}. The simplest 
    SIE model introduced a phenomenological parameter $f_e$ to account for any possible difference 
    between the true velocity dispersion and that of the SIS model: 
    \begin{align}
        \theta_E = 4\pi \frac{d_A (z_l, z_s)}{d_A (0, z_s)} \left( \frac{f_e \sigma}{c} \right)^2
    \end{align}
    but, as mentioned in \S~\ref{sec:datasets}, the optimal value of $f_e$ is very close to $1$ 
    so it does not provide any effective variation away from a pure SIS.
    
    To understand the influence of the lensing galaxy's mass distribution on our CDD test, we 
    therefore also consider an extended SIE model based on an assumed power-law density profile $\rho (r)$ and 
    a luminosity density of stars $\nu (r)$ \citep{koopmans2005proc}:
    \begin{align}
        \rho (r) &= \rho_0 \left( \frac{r}{r_0} \right)^{-\tau} \ , \\
        \nu (r) &= \nu_0 \left( \frac{r}{r_0} \right)^{-\gamma} \ ,
    \end{align}
    where $r$ is the spherical radial coordinate from the center of the lensing galaxy; and $\tau$ and 
    $\gamma$ are adjustable free parameters. The observed velocity dispersion $\sigma_0$ can provide a 
    dynamical estimate of the mass, based on this power-law density profile. The corresponding central 
    velocity disperion of the extended-SIE model is \citep{doi:10.1093/mnras/stw932} 
    \begin{align}
        \sigma_{\text{ex-SIE}}^2 =& \left( \frac{c^2}{4} \frac{d_A (0, z_s)}{d_A (z_l, z_s)} \theta_E \right) \frac{2}{\sqrt{\pi} (\xi - 2\beta)} \left( \frac{\theta_{\mathrm{ap}}}{\theta_E} \right)^{2-\tau} \notag \\
        & \times \left[ \frac{\lambda (\xi) - \beta \, \lambda (\xi + 2)}{\lambda (\tau) \, \lambda (\gamma)} \right] \frac{\Gamma (3 - \xi/2)}{\Gamma (3 - \gamma/2)} \ ,
    \end{align}
    where 
    \begin{itemize}
	\item $\beta$ characterizes the anisotropic distribution of the three-dimensional velocity dispersion, which
        appears to be Gaussian with $\beta = 0.18 \pm 0.13$, based on a sample of local elliptical galaxies 
        \citep{1538-3881-121-4-1936};
	\item $\theta_{\mathrm{ap}}$ is the spectrometer aperture radius;
	\item $\xi \equiv \tau + \gamma - 2$;
	\item $\lambda (x) \equiv \Gamma \left(\frac{x-1}{2} \right) \Big/ \Gamma \left( \frac{x}{2} \right)$ is the ratio of Euler's gamma functions.
    \end{itemize}
    
    We use the distances $d_A (0, z_s)$ and $d_A (z_l, z_s)$ to extract $\sigma$ from the above equation, with
    the help of the CDD relation $d_A (z) = {d_L(z) \, \eta(z)}/{(1 + z)^2}$ and the HII galaxy data. The result is
    \begin{align}
        \frac{d_A (0, z_s)}{d_A (z_l, z_s)} &= \frac{1}{d_{\mathrm{ratio}}} = \left[ 1 - \frac{1+z_l}{1+z_s} \frac{d_A (z_l)}{d_A (z_s)} \right]^{-1} \\
        &= \left[ 1 - \frac{1+z_s}{1+z_l} \frac{\eta (z_l)}{\eta (z_s)} \frac{d_L (z_l)}{d_L (z_s)} \right]^{-1} \\
        &= \left[ 1 - \frac{1+z_s}{1+z_l} \frac{\eta (z_l)}{\eta (z_s)} 10^{ (\mu^{\mathrm{obs}} (z_l) - \mu^{\mathrm{obs}} (z_s) ) / 5 } \right]^{-1} \ .
    \end{align}
    But notice that we now have two more free parameters (i.e., $\tau$ and $\gamma$) in the extended SIE model, 
    than we did with the simple SIS. The overall number of adjustable variables is now large enough to create
    degeneracy in the optimization. As such, we restrict our attention solely to the formulation 
    $\eta (z) = 1 + \eta_0 z$ (i.e., Eq.~20) to mitigate this excessive flexibility. 

    Our optimization strategy now relies on identifying the best-fit values of $\tau$ and $\gamma$, for which
    the relevant $\chi^2$ function is 
    \begin{align}
        \chi^2 = \sum_{i=1}^{53} \left[ \frac{ \sigma^{\text{ex-SIE}}_i (\tau, \gamma, \beta; \eta_0) - \sigma_{i}^{\mathrm{obs}} }{ \sigma_{\sigma_{0,i}} } \right]^2 \ ,
    \end{align}
where the sum is taken over the 53 strong lens systems identified in Table~1 above.

\begin{figure}
    \plotone{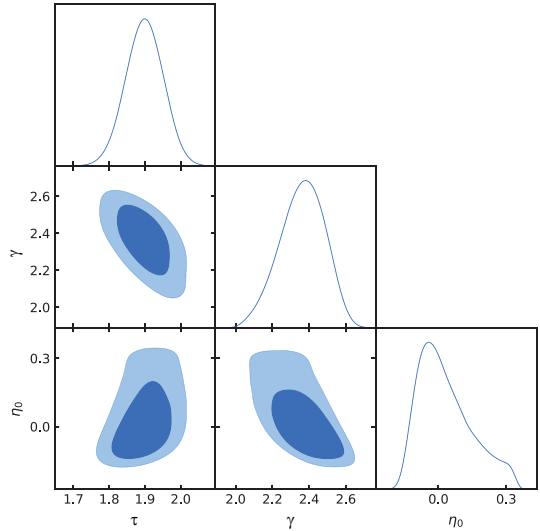}
    \caption{Posterior probability density function of the parameter $\eta_0$, and the power-law lens model
    indices $\tau$ and $\gamma$. \label{fig:exSIE0}}
    \end{figure}

The 1D and 2D marginalized distribution plots produced with this procedure are shown in fig.~\ref{fig:exSIE0} for the 
CDD parameter $\eta_0$ and the power-law indices $\tau$ and $\gamma$. The corresponding $68\%$ confidence limits are
    \begin{align}
        \eta_0 &= 0.0093^{+0.1520}_{-0.0939} \ , \\
        \tau &= 1.8990^{+0.0521}_{-0.0526} \ , \\
        \gamma &= 2.3680^{+0.1185}_{-0.1305} \ .
    \end{align}
    
    Our optimized values of $\tau$ and $\gamma$ are consistent with those found in previous work \cite{0004-637X-834-1-75}, 
    in which the reported results were $\tau= 1.97 \pm 0.04$ and $\gamma = 2.40 \pm 0.13$. We note that \citet{0004-637X-834-1-75} 
    used the model dependent Type Ia SN data with the CDD to extract the angular-diameter distance, whereas we have used the
    HII galaxy Hubble diagram, yet the use of an optimized power-law lens model has not changed the outcome significantly. 
    The best-fit value of $\eta_0$ is still fully consistent with the CDD, but the confidence limit is much wider than that 
    of a pure SIS model (compare Eqs.~24 and 34). This slight weakening of the CDD constraint is entirely due to the
    larger number of free parameters in the extended SIE model. We conclude from this comparison that the use of a more
    elaborate SIE lens model (such as the extended power-law profile we have examined here) will probably not improve the 
    results we have obtained with a simple SIS, in spite of the lingering scatter associated with this simplified mass
    profile (see figs.~2 and 3). A more direct empirical determination of the mass distribution within the lenses will
    be required to significantly refine the results reported here. 
    
    \section{Conclusion}\label{sec:conc}
    A commonly used method of testing the CDD has been to compare the luminosity 
    distance derived from Type Ia SNe with the angular-diameter distance measured 
    using galaxy clusters. But these are no longer the only standard candles and 
    rulers available today. Seeking alternatives is desirable because the older approach
    requires the pre-assumption of specific cosmological models in order to optimize
    the `nuisance' parameters in the distance-redshift relation. This limiting factor
    can lead to additional uncertainty and bias, which may explain why some earlier 
    work with the CDD has produced conflicting results. A summary of previous 
    inconsistencies may be found in \citet{melia2018model}. 
    
    In this paper, we have chosen a new combination of standard candles (HII galaxies)
    and rulers (strong lenses) to test the CDD without the need to pre-assume any
    particular cosmology. The fact that our analysis shows consistency with
    the CDD at a very high level of confidence is therefore quite compelling because 
    we have avoided introducing unknown systematics associated with particular models. 
    We do note, however, that we have assumed spatial flatness throughout our analysis,
    which appears to be consistent with a broad range of cosmological measurements.
    Nevertheless, this is a caveat to keep in mind, should any new evidence emerge
    that the Universe is not spatially flat. 
    
    Another important benefit of our
    model-independent study is that the additional flexibility in fitting the data
    otherwise present when cosmology-dependent parameterizations are introduced
    is absent from our approach. Our test is therefore straightforward and clean
    because any possible variations in $\eta(z)$ away from $1$ cannot be attributed 
    to the cosmology itself. Were we to find that $\eta(z)\neq 1$, the evidence
    in favor of new physics would therefore have been stronger with our method 
    than what would be found using model-dependent data.

    The results in this paper fully confirm another recent model-independent test
    of the CDD carried out by \citet{melia2018model}. In that work, the standard
    ruler was provided by compact quasar cores. We are therefore starting to see
    a consistent pattern of results in which model-independent tests all agree 
    that the CDD is realized in nature. 

    The principal caveat of our work is the irreducible scatter in our
    CDD data stemming from the use of a simplified single isothermal sphere (SIS) 
    model for the lens. We have attempted to mitigate the impact of an imprecisely
    known mass distribution in the lens by also considering an extended SIE model, 
    in which the internal structure of the lens is characterized by power laws
    for the mass and luminosity densities, with two adjustable indices. While
    this allows greater freedom in modeling the lens itself, however, the downside with 
    such an approach is the additional degeneracy offered by the greater flexibility
    with the overall optimization of the parameters. Our results for both the simple
    SIS and the extended SIE lens models confirm the CDD all the way out to $z\sim 2.3$,
    with a violation no bigger than $\eta_0\sim 0.01-0.015$, in a parameterization
    $\eta(z)=1+\eta_0z$. But the CDD constraint is actually weaker with the 
    extended SIE lens (i.e., $\eta_0=0.0093^{+0.1520}_{-0.0939}$ versus
    $\eta_0=0.0147^{+0.056}_{-0.066}$) due to the less precise confidence range of 
    the best fit values resulting from the adjustable power-law distributions. 

    In future work with strong lensing, it would be desirable to determine the lens 
    mass distribution more accurately,
    e.g., using ray tracing, rather than simply relying on a SIS model, which 
    appears to produce irreducible scatter in the results due to its over-simplification
    of the lens structure. Eventually, this improvement should yield a better measurement 
    of the angular-diameter distance, allowing us to test the CDD with even higher 
    precision than is available today.
    
    \section*{Acknowledgements}
We are grateful to the anonymous referee for recommending several
improvements to this manuscript. This work was supported by the National Key R \& D
Program of China (2017YFA0402600), the National Science Foundation of China (Grants
No.11573006, 11528306), the Fundamental Research Funds for the Central Universities
and the Special Program for Applied Research on Super Computation of the NSFC-Guangdong
Joint Fund (the second phase).

    \listofchanges
    \end{document}